\documentclass[aps,prd,superscriptaddress,preprintnumbers,floatfix]{revtex4}
\usepackage{graphicx,amsmath,amssymb,dcolumn,bm}
\usepackage{fixmath}
\usepackage{feynmp}
\usepackage{epic,eepic}
\usepackage{wrapfig}
\usepackage{slashed}
\allowdisplaybreaks[2]
\usepackage{calrsfs}
\DeclareMathAlphabet{\pazocal}{OMS}{zplm}{m}{n}

\begin{document}
\preprint{KEK-TH-2388, J-PARC-TH-0265}
\title{J-PARC hadron physics and future possibilities on color transparency}
\author{S. Kumano}
\affiliation{KEK Theory Center,
             IPNS,
             KEK,
             Oho 1-1, Tsukuba, Ibaraki, 305-0801, Japan}
\affiliation{J-PARC Branch, KEK Theory Center,
           IPNS, KEK,\\
           and Theory Group, Particle and Nuclear Physics Division, 
           J-PARC Center, \\
           Shirakata 203-1, Tokai, Ibaraki, 319-1106, Japan}
\date{March 31, 2022}

\begin{abstract}
The J-PARC is a hadron-accelerator facility to provide secondary beams 
of kaons, pions, neutrinos, muons, and the others together 
with the primary proton beam for investigating a wide range 
of science projects. High-energy hadron physics can be studied by
using high-momentum beams of unseparated hadrons, which are essentially pions, 
and also primary protons. In this report, possible experiments are explained 
on color transparency and generalized parton distributions (GPDs).
These projects are complementary to lepton-scattering experiments 
at JLab, COMPASS/AMBER, and future electron-ion colliders.
Because of hadron-beam energies up to 30 GeV, the J-PARC is a unique facility 
to investigate the transition region from the hadron degrees of freedom 
to the quark-gluon one.
It is suitable for finding mechanisms of the color transparency. 
Such color-transparency studies are also valuable 
for clarifying factorization of hadron-production processes
in extracting the GPDs from actual measurements.
These studies will lead to the understanding of basic high-energy hadron
interactions in nuclear medium and to clarifications
on the origins of hadron spins, masses, and internal pressure mechanisms.\\

\end{abstract}
\maketitle

\section{Introduction}

The Japan Proton Accelerator Research Complex (J-PARC) is a multi-purpose 
facility ranging from life sciences to nuclear and particle physics
\cite{J-PARC-web}.
By using secondary beams of kaons, pions, neutrinos, muons, 
and the others as well as the primary proton beam, 
a wide range of experiments are possible.
A well-known particle physics experiment is on neutrino oscillations,
and there are also promising projects, including future ones,
to probe physics beyond the standard model such as 
by lepton-flavor violation, $g-2$,
rare kaon decays, and the neutron electric-dipole moment. 
It is a flagship facility not only in particle physics 
but also in hadron physics.

The J-PARC hadron physics, so far, focuses on low-energy hadron
and nuclear physics with strangeness. It is a unique project 
in extending our knowledge of standard nuclear physics 
by exploring the new flavor degrees of freedom, strangeness.
The recent highlights are discoveries on 
(1) event excess observed by the missing spectrum of $^{12} C (K^-,p)$
toward understanding kaonic- and hyper-nuclei,
(2) the $\bar K NN$ bound state in $^3 He(K^- , \Lambda p) n$,
(3) $\Xi^- $--$ ^{14}N$ bound system with the $\Xi^-$ binding energy 
    of 1.27$\pm$0.21 MeV and subsequent findings on other bound states 
    \cite{J-PARC-recent-highlights}.
Such strangeness nuclear-physics studies will be continued at J-PARC,
and we will expand our knowledge of the current nuclear physics
to the new realm.
These investigations will lead to the understanding of the equation of state
for neutron-star matter, which is a developing field with recent 
progress on astrophysical observations including gravitational waves.
Right now, the hadron-hall extension is under consideration,
and charmed hadron physics will be also an interesting topic at J-PARC.

In addition, high-momentum beamlines with the primary 30-GeV protons
and unseparated hadrons, mainly pions, are now available up to
about 20-GeV/$c$ momentum. These hadron beam energies are not very high,
which indicates a unique opportunity to investigate 
the intermediate region from hadron degrees of freedom (d.o.f.)
to the quark and gluon d.o.f. described by perturbative QCD.
In this sense, the J-PARC is a similar facility to 
the Thomas Jefferson National Accelerator Facility (JLab);
however, it is complementary to JLab projects
because different aspects can be investigated 
by hadron reactions. 

Such complementarity is clear in the studies of the color transparency
and the generalized parton distributions (GPDs).
The color transparency means that a hadron is expected to 
pass freely through the nuclear medium at large momentum transfer.
There was a mysterious experimental result from 
the BNL-AVA collaboration in 2004 that the transparency increases
with the beam momentum as expected; however, it started to 
drop at about 10 GeV \cite{BNL-EVA-2004}. 
It is inconsistent with the color transparency
prediction, which cannot be explained at this stage. 
Since the BNL measurement is not understood easily, it should be confirmed
by independent experiments and the J-PARC is the ideal facility
for this experiment. There was a new color-transparency measurement
from the JLab that the transparency does not appear even up to
$Q^2=$14 GeV$^2$ in 2021 \cite{jlab-trans-2021}. 
Together with this new development, a possible J-PARC experiment 
becomes more valuable for finding the mechanism.
It is a timely project by considering the new experimental 
development at JLab in connection with the peculiar BNL result
and because the high-momentum beam line is now ready at J-PARC.

On the other hand, the GPDs can be investigated in the same
processes at hadron accelerator facilities
\cite{KSS-GPD-2009,J-PARC-GPD-2016,J-PARC-GPD-LoI}. 
In particular, the color transparency argument is essential
for factorization of the processes in extracting the GPDs
from experimental measurements \cite{Collins:1997hv}.
The GPDs are valuable quantities for understanding the origins
of hadron mass and spin compositions \cite{GPD-review}. 
They contain information on form factors and 
parton distribution functions (PDFs),
and their second moments are related to
quark orbital-angular-momentum contributions
to the nucleon spin. Furthermore, they are related
to gravitational form factors of hadrons as shown,
for example, in Ref.\,\cite{KSS-grav-2018},
they are also important for finding mechanisms of hadron mass
and internal pressure generations.
Therefore, the GPDs will play a crucial role in future developments
of hadron physics.
We believe that their studies will lead 
to the understandings on the origins of nucleon spin and mass
in terms of quarks and gluons.
In this report, we explain possible high-energy hadron-structure physics
especially related to color transparency, which is the main topic of
this workshop, along with possible projects on the GPDs.

\section{Transition from hadron to quark-gluon degrees of freedom
         and color transparency}

\subsection{Hard exclusive reactions and constituent counting rule}

\begin{wrapfigure}[33]{r}{0.42\textwidth}
   \vspace{-0.80cm}
   \begin{center}
     \includegraphics[width=5.0cm]{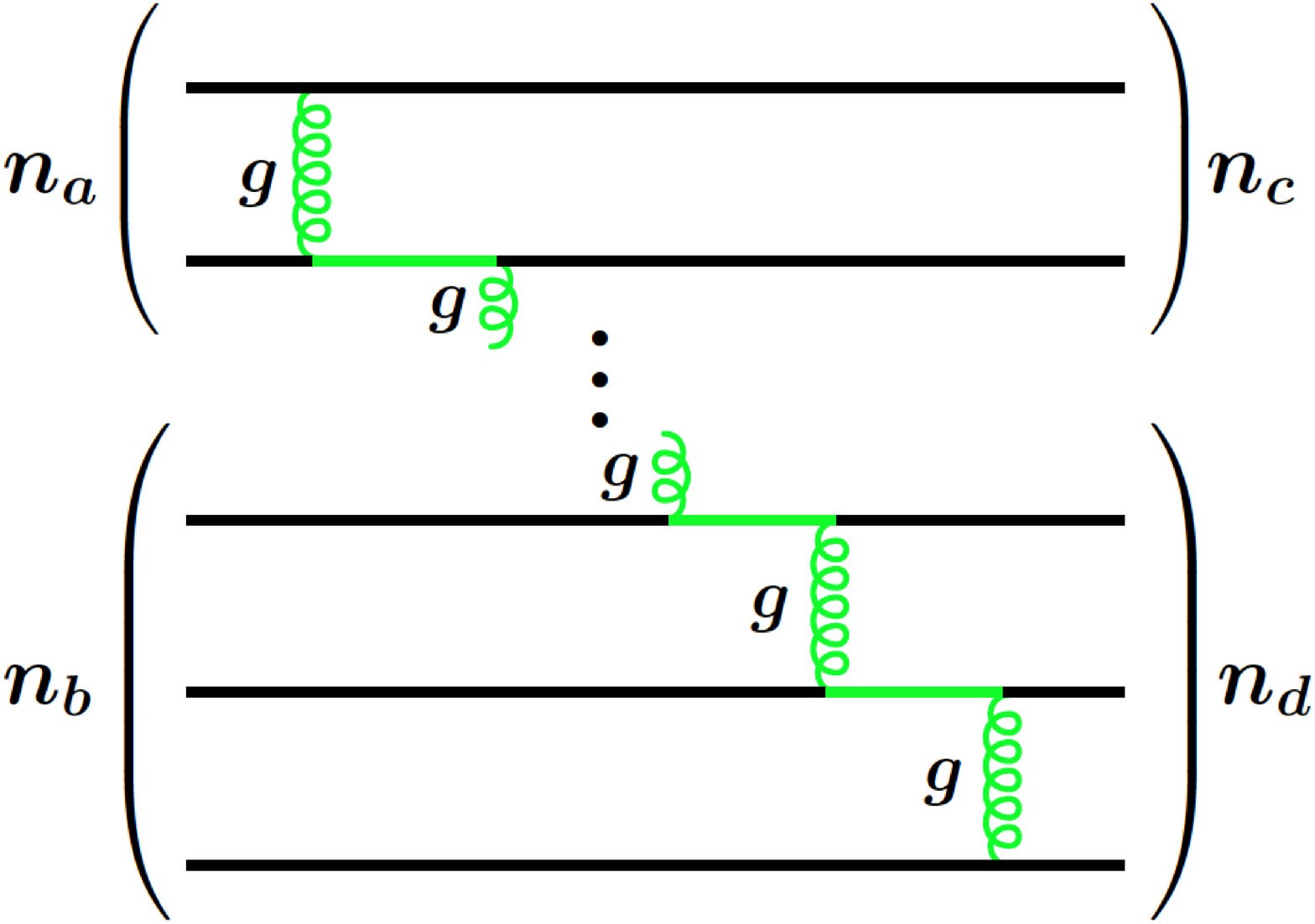}
   \end{center}
\vspace{-0.80cm}
\caption{
Hard-gluon exchange process
in an exclusive hadron reaction.
}
\label{fig:hard-glun-exchange}
\vspace{0.10cm}
\begin{center}
     \includegraphics[width=5.5cm]{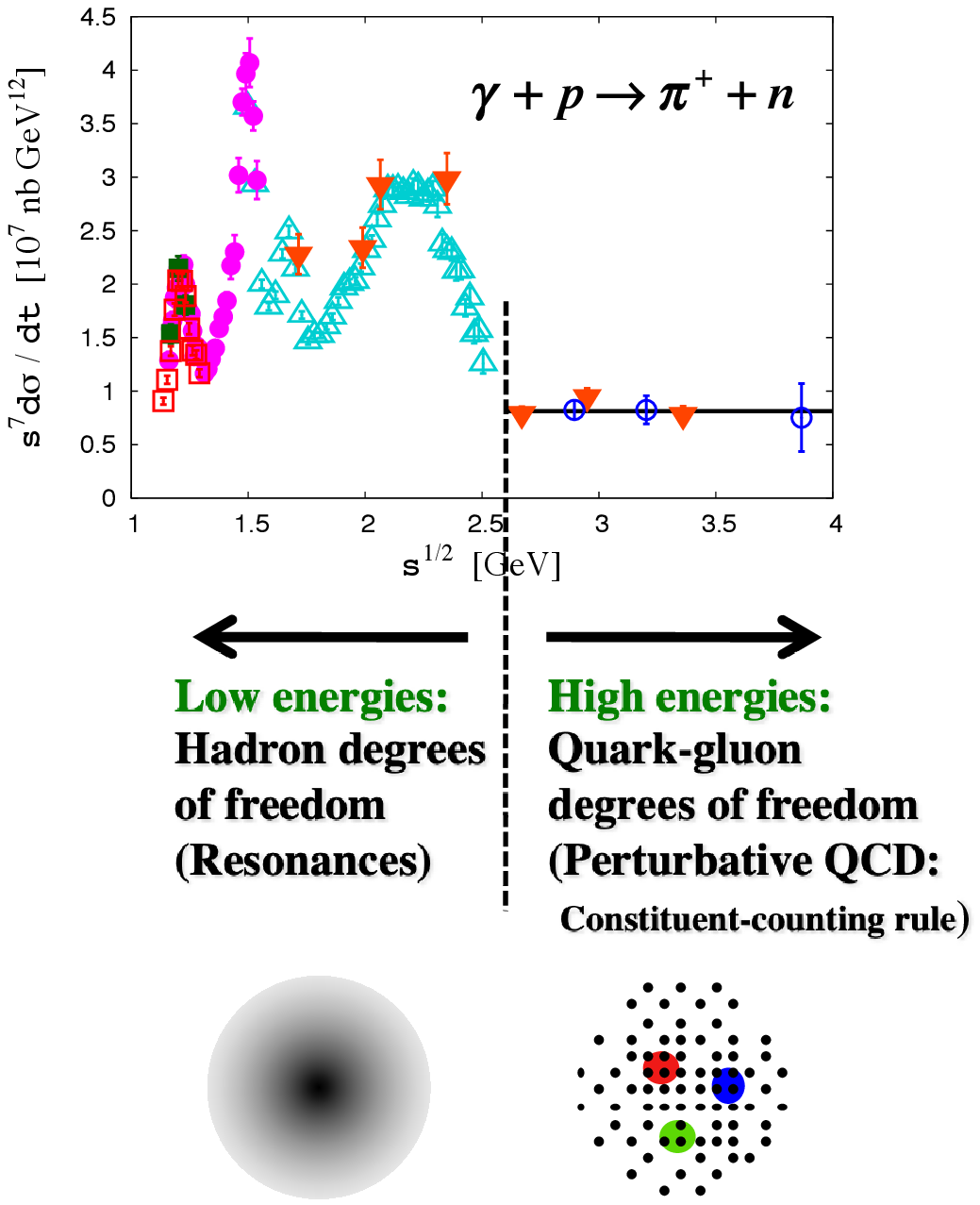}
\end{center}
\vspace{-0.75cm}
\caption{
Transition from hadron d.o.f.
to quark-gluon d.o.f. and constituent
counting rule in $\gamma + p \to \pi^+ + n$.
}
\label{fig:transition-counting}
\vspace{-0.5cm}
\end{wrapfigure}

Let us consider a high-energy exclusive reaction.
The cross section for a hard exclusive hadron reaction $a+b \to c+d$ 
is generally expressed by the partonic scattering term $H_{ab\to cd}$ 
and lightcone amplitudes $\phi_h  \ (h=a,\,b,\,c,\,d)$, 
and its matrix element is given by
\cite{kks-2013,kk-2014,cks-2016}
\vspace{-0.20cm}
\begin{align}
& M_{ab \to cd} =  \int [dx_a] \, [dx_b] \, [dx_c] \, [dx_d]  \,
    \phi_c ([x_c]) \, \phi_d ([x_d]) 
\nonumber \\[-0.1cm]
& 
\times 
H_{ab \to cd} ([x_a],[x_b],[x_c],[x_d],Q^2) \, 
      \phi_a ([x_a]) \, \phi_b ([x_b]) .
\label{eqn:mab-cd}
\end{align}
\ \vspace{-0.5cm}

\noindent
Here, $[x_h]$ is a set of the lightcone momentum fractions $x_i=p_i^+/p_h^+$ 
for the parton $i$ and the hadron $h$.
The matrix element or the cross section is described by
the hard part $H_{ab\to cd}$ calculated in perturbative QCD 
and the soft one given by the distribution amplitudes.
A typical hard exclusive process is shown in Fig.\,\ref{fig:hard-glun-exchange}.
The exclusive reaction is possible if the hard momentum is shared
among the constituents, which should be close to each other 
(``small hadron component") to form a hadron in the final state.
Assigning the hard gluon and quark propagators as well as external
quark factors, we obtain the cross section expressed by
the Mandelstam variables $s$ and $t$ as
\cite{kks-2013,kk-2014,cks-2016,ss-2011,abrs-2021}
\vspace{-0.20cm}
\begin{align}
\frac{d\sigma_{ab \to cd}}{dt} = \frac{1}{s^{\, n-2}} \, f_{ab \to cd}(t/s),
\label{eqn:cross-counting}
\end{align}
\ \vspace{-0.5cm}

\noindent
where $f_{ab \to cd}(t/s)$ is a function which depends on the scattering
angle, and $n$ is the total number of constituents 
($n = n_a+n_b+n_c+n_d$) involved in the reaction.

This constituent-counting rule was confirmed by high-energy 
exclusive reactions. For example, the cross-section 
($s^7 d\sigma/dt$) data are shown for $\gamma +p \to \pi^+ +n$ 
\cite{JLab-counting-2003}
as the function of the center-of-mass energy $\sqrt {s}$ 
in Fig.\,\ref{fig:transition-counting}.
At low energies, it indicates bumpy resonance structure,
and it becomes constant at high energies, 
as predicted by the constituent counting rule.
Therefore, hadrons and nuclei should be 
described by quark and gluon degrees of freedom 
in high-energy reactions. However, the transition energy depends
on the reaction process and it is not easily determined.
There are two major reasons for this issue.
One is that there are too many processes like 
Fig.\,\ref{fig:hard-glun-exchange},
and the other is that the distribution amplitudes are
not well understood except for the pion one.

\subsection{Color transparency}

\begin{figure}[h!]
\vspace{-0.50cm}
\hspace{0.0cm}
\begin{minipage}[c]{0.95\textwidth}
    \vspace{+0.00cm}\hspace{-0.00cm}
    \begin{center}
     \includegraphics[width=12.0cm]{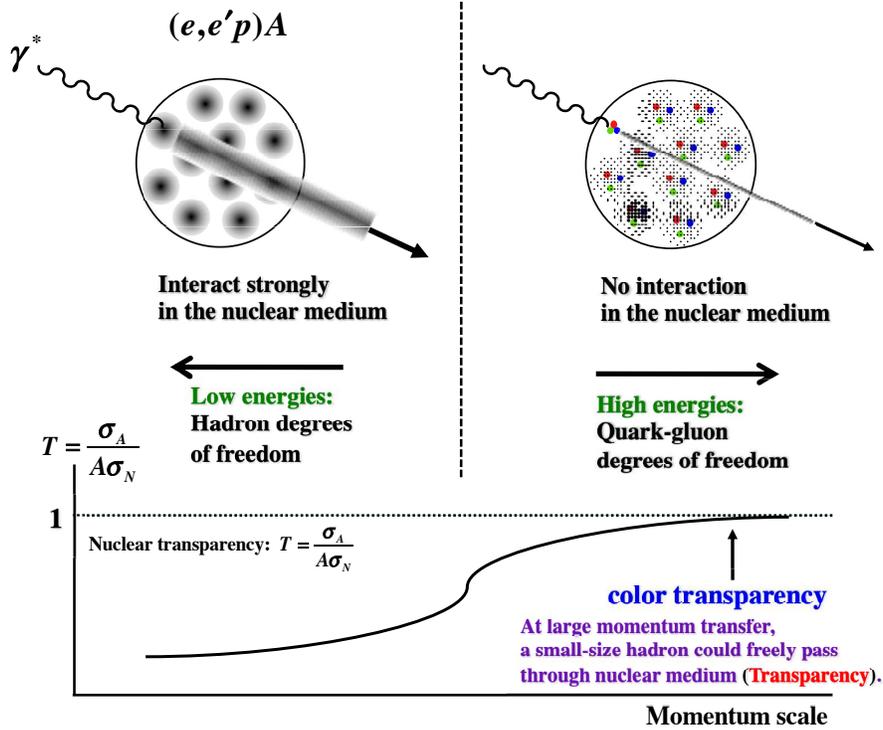}
    \end{center}
\vspace{-0.70cm}
\caption{Color transparency in high-energy electron scattering $(e,e' p)A$.}
\label{fig:asym-ex-ey}
\end{minipage}
\vspace{-0.10cm}
\end{figure}

Hadron interactions in nuclear medium are valuable
for understanding dynamical properties of QCD
and for application to high-energy hadron reactions.
As seen in the constituent-counting rule in the exclusive hadron reactions,
the internal constituents should stick together to form a hadron.
Namely, a small hadron component dominates
the reaction at a large-momentum transfer.
The small-size hadron is expected to pass through 
the nuclear medium without interactions.
This phenomena is called color transparency, and 
the nuclear transparency $T=\sigma_A/(A \sigma_N)$
is used for showing this quantity.
Here, $\sigma_A$ and $\sigma_N$ are nuclear and nucleon
cross sections, respectively, and $A$ is the mass number of the nucleus.                                                
The nuclear transparencies are measured by the reactions
$e A \rightarrow e' p (A-1)$ and
$p A \rightarrow 2p (A-1)$
at electron and proton accelerator facilities.
The color transparency involves both the energy scale 
at which a hadron is formed in a point-like configuration 
as discussed in Sec. 2.1, and how it then evolves through the medium.
For example, $A(p,2p)$ measurements were done at about
90$^\circ$ in the $pp$ center-of-mass frame
in the BNL-AGS experiment \cite{BNL-EVA-2004},
so that the large-momentum-transfer mechanism of
Sec. 2.1 could be applied. On the other hand, 
the expansion of the wave packet is controlled 
by the proton momentum in the final state
as its expansion is characterized by the coherence length
which is proportional to the proton momentum \cite{2010-ks}.
The color transparency contains these two mechanisms.
The $(e,e' p) A$ reaction is illustrated 
in Fig.\,\ref{fig:asym-ex-ey} together with our expectation 
on the nuclear transparency.
At low energies, hadrons interact strongly with nucleons 
in the nucleus, so that the nuclear transparency is small.
As the energy increases, it becomes larger and $T$ should 
become one in the high-energy limit according to the color 
transparency.

\begin{wrapfigure}[12]{r}{0.42\textwidth}
   \vspace{-0.60cm}
   \begin{center}
     \includegraphics[width=4.5cm]{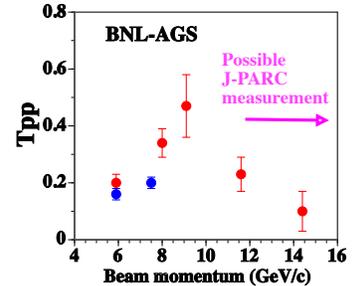}
   \end{center}
\vspace{-0.80cm}
\caption{BNL experiment on nuclear
transparency by $p A \rightarrow 2 p (A-1)$.}
\label{fig:BNL-experiment}
\vspace{-0.5cm}
\end{wrapfigure}

The BNL-AGS data \cite{BNL-EVA-2004}
are shown in Fig.\,\ref{fig:BNL-experiment}
for the nuclear transparency in the reaction 
$p A \rightarrow 2 p (A-1)$ as the function of the beam momentum. 
As the hard scale of the reaction, the proton-beam energy in this case,
becomes larger, the transparency is expected to increase.
In fact, the transparency increases as the momentum becomes larger up
to $p \sim 10$ GeV/$c$; however, it drops at $p > 10$ GeV/$c$.
This result seems to be in contradiction to the color transparency
expectation and it is not easily explained theoretically.
Since the J-PARC has the 30-GeV primary-proton-beam energy,
the larger momentum region can be measured and 
the transparency phenomena shown in Figs.\,\ref{fig:asym-ex-ey}
and \ref{fig:BNL-experiment} should be tested. Therefore, the J-PARC 
provides a valuable opportunity for investigating basic 
hadron interactions in the nuclear medium as predicted by perturbative QCD.

\begin{figure}[h!]
\vspace{-0.20cm}
\begin{minipage}{0.51\textwidth}
\vspace{0.60cm}
   \begin{center}
     \includegraphics[width=4.5cm]{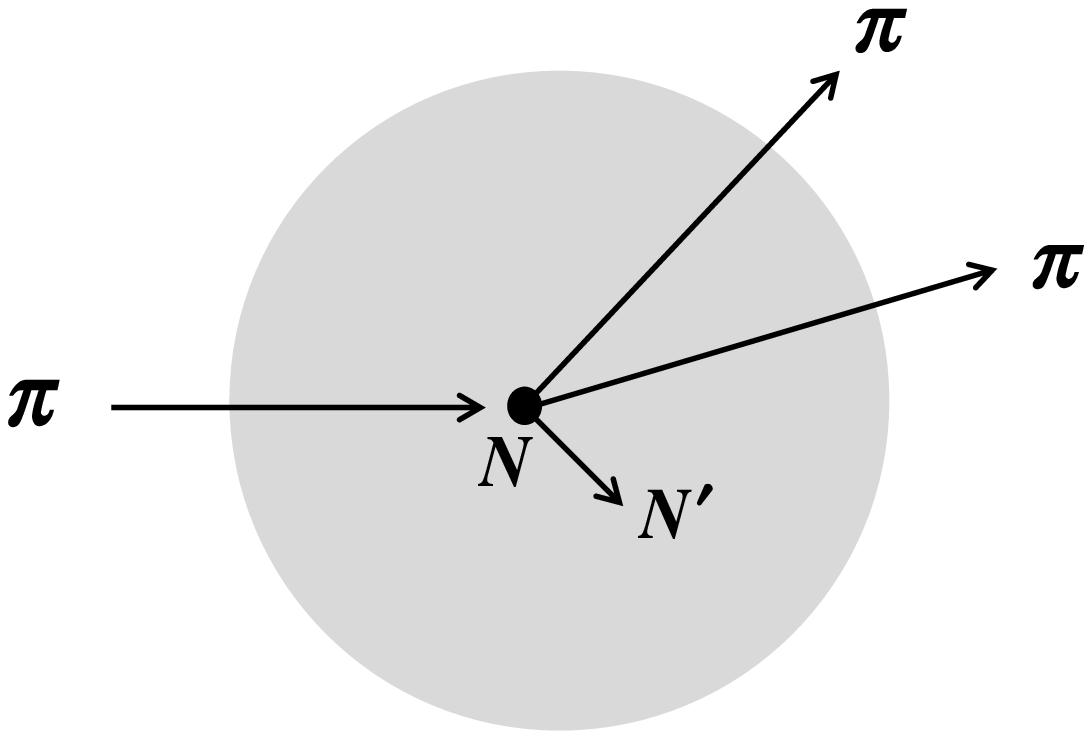}
   \end{center}
\vspace{-0.40cm}
\caption{$\pi^- A \to \pi^- \pi^+ A^*$ for color transparency.}
\label{fig:pion-2-3-color}
\vspace{0.20cm}
\end{minipage}
\hspace{-1.5cm}
\begin{minipage}{0.48\textwidth}
   \begin{center}
     \includegraphics[width=5.3cm]{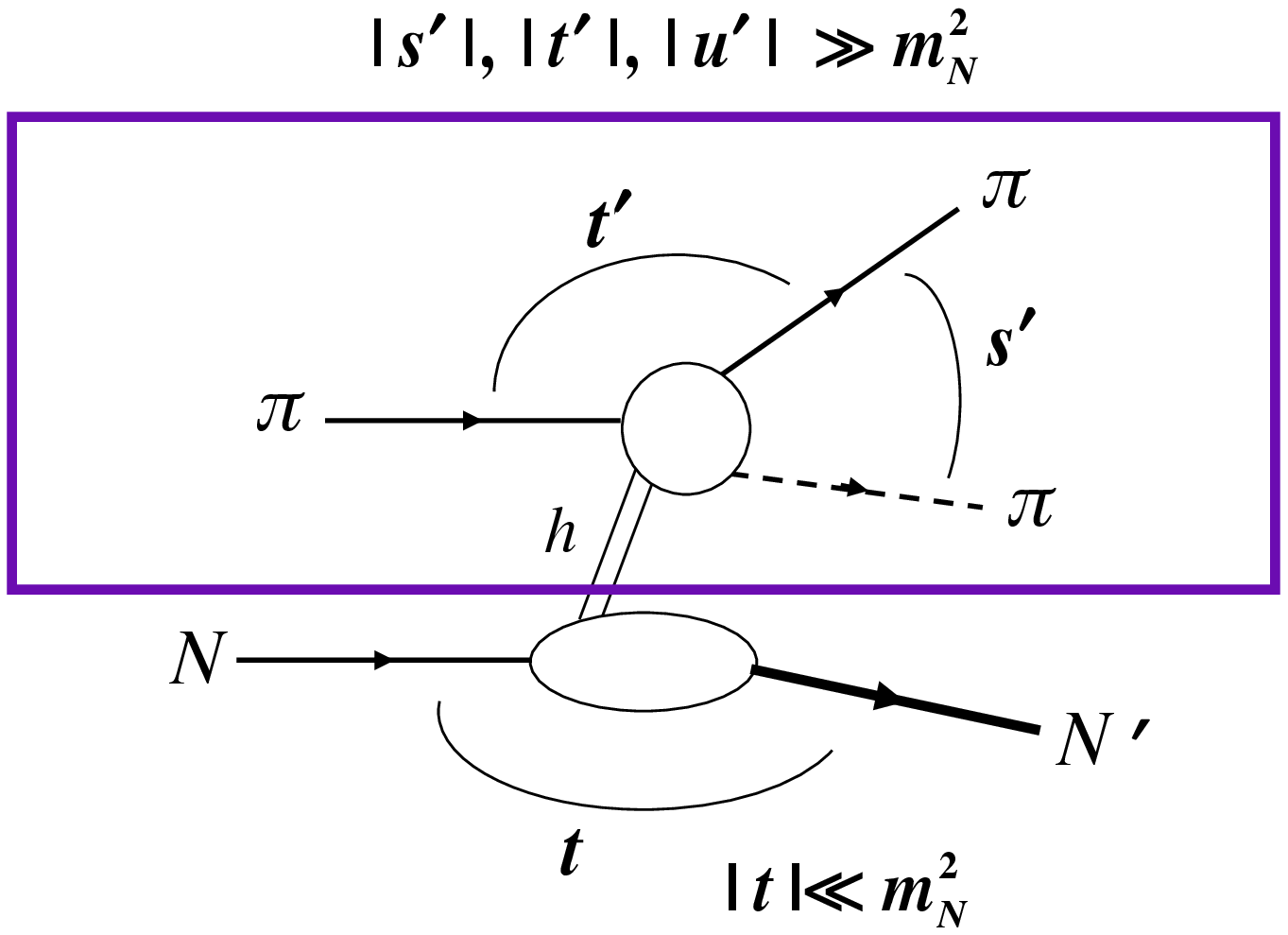}
   \end{center}
\vspace{-0.80cm}
\caption{$\pi^- N \to \pi^- \pi^+ p$ process.}
\label{fig:pion-2-3-process}
\end{minipage} 
\vspace{-0.30cm}
\end{figure}

Next, we proposed a new $2 \to 3$ reaction process
for investigating the color transparency phenomena by
using the high-momentum pion beam at J-PARC and COMPASS /AMBER 
\cite{2010-ks}.
In particular, we studied hard branching $2\to 3$ processes 
$\pi^- A \to \pi^- \pi^+ A^*$ as illustrated 
in Fig.\,\ref{fig:pion-2-3-color}.
The basic reaction process $\pi^- N \to \pi^- \pi^+ N'$ is shown
in Fig.\,\ref{fig:pion-2-3-process}, where the kinematics of
the upper hard-reaction part $\pi h \to \pi \pi$ is controlled
by the lower part $N \to h N'$.
This hard branching $2\to 3$ processes 
with nuclei provide an effective way to determine 
the momentum transfers needed for effects of point-like 
configurations to dominate large-angle $2\to 2$ processes.

\begin{wrapfigure}[10]{r}{0.36\textwidth}
   \vspace{-0.60cm}
   \hspace{0.20cm}
   \begin{center}
     \includegraphics[width=4.8cm]{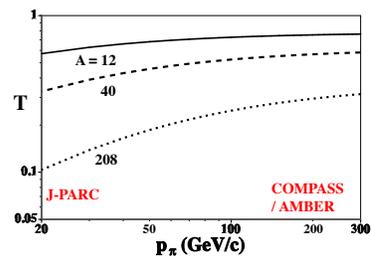}
   \end{center}
\vspace{-0.85cm}
\caption{Color transparency in $2 \to 3$
process $\pi^- A \to \pi^- \pi^+ A^*$.} 
\label{fig:pion-experiment}
\vspace{-0.5cm}
\end{wrapfigure}

Calculated nuclear transparencies are
shown in Fig.\,\ref{fig:pion-experiment}
as the functions of the pion-beam energy for three different nuclei
\cite{2010-ks}.
The $2\to 3$ reaction allows the effects of the transverse size 
of configurations to be decoupled from effects of the space-time
evolution of these configurations, which is different from
past $2\to 2$ methods.
Namely, it is a new method for probing the dynamics
of large angle hadron-hadron scattering using the color-transparency 
phenomenon which is free from the limitations imposed 
by the expansion effects of the point-like configurations.
The beam energies are in the 20-200 GeV range,
namely from the J-PARC to COMPASS/AMBER energies,
and they are appropriate for these $2 \to 3$ reaction studies. 

\section{Possible J-PARC project on generalized parton distributions}
\vspace{-0.00cm}

\subsection{J-PARC hadron facility}
\vspace{-0.05cm}

The J-PARC provides the most intense proton beam 
in the energy region above multi-GeV, 
and it is a multi-purpose facility from 
nuclear and particle physics, material and life sciences 
to industrial applications.
Nuclear and particle physics experiments are done
by secondary beams (pions, kaons, neutrinos, muons, and antiprotons)
as well as the primary 30-GeV proton beam.

\begin{wrapfigure}[11]{r}{0.40\textwidth}
   \vspace{-0.65cm}
   \begin{center}
\includegraphics[width=5.5cm]{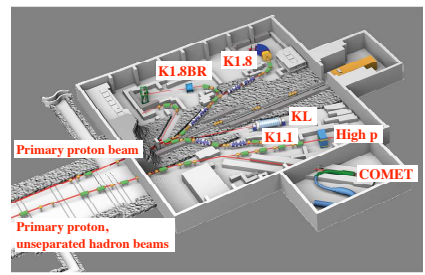}
   \end{center}
\vspace{-0.70cm}
\caption{J-PARC hadron hall.}
\label{fig:hadron-hall}
\vspace{-0.5cm}
\end{wrapfigure}

Hadron-physics projects are done in the hadron experimental 
hall of Fig.\,\ref{fig:hadron-hall}.
There are also particle-physics projects on
lepton-flavor violation and rare kaon decays 
in this hadron hall.
The K1.8 indicates kaons with momentum around 1.8 GeV/c 
and this beamline is used for strangeness nuclear physics
such as strangeness $-$2 hypernuclei with $\Xi^-$ by
the $(K^-,K^+)$ reaction. The K1.1 beamline is intended 
for low-momentum stopped kaon experiments such as the studies 
of kaonic nuclei.
The high-momentum beamline indicates for 30 GeV protons 
and unseparated hadrons, which are essentially pions
with up to about 20 GeV/$c$ momentum.
Using these high-momentum protons and pions, we can
investigate interesting high-energy hadron physics topics,
particularly in the transition region from hadron 
to quark-gluon degrees of freedom, as explained
in this paper.

\vspace{-0.15cm}
\subsection{Generalized parton distributions}
\vspace{-0.05cm}

The studies of the color transparency are closely related
to the GPD measurements in connection with factorization 
and reaction mechanisms.
The spacelike GPDs have been measured in deeply virtual Compton scattering
and meson productions at charged-lepton accelerator facilities.
On the other hand, the timelike GPDs are measured in two-photon processes
of $e^+ e^-$ reactions, for example, at the KEK-B facility
\cite{KSS-grav-2018}.
Furthermore, a GPD measurement is also considered 
at the Fermilab neutrino facility \cite{GPD-neutrino}.
In addition to these projects, it is possible to investigate the GPDs
at hadron accelerator facilities such as the J-PARC.

\begin{wrapfigure}[8]{r}{0.31\textwidth}
   \vspace{-0.85cm}
    \begin{center}
     \includegraphics[width=4.4cm]{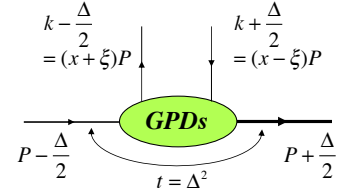}
    \end{center}
\vspace{-0.75cm}
\caption{GPD kinematics.}
\label{fig:GPD-kinematics}
\end{wrapfigure}

The quark GPD is given by the amplitude to extract a quark
from a hadron and then to insert it into the hadron
at a different spacetime point as shown in Fig.\,\ref{fig:GPD-kinematics}.
The generalized scaling variable $x$ and a skewdness parameter $\xi$
are defined by the lightcone momenta $k^+$ and $P^+$ as
$x = k^+ / P^+$ and $\xi = - \Delta^+ /(2 P^+)$.
Here, $k$ is the average momentum of the quarks,
and $\Delta^{\mu}$ is the four momentum transfer $\Delta = p' - p$
for the hadron or the quark.
There is another variable $t$ defined by $t=\Delta^2$.
The variable $x$ indicates the momentum fraction carried by a quark
in the hadron, and the skewdness parameter $\xi$ 
indicates the momentum-transfer fraction.
The GPDs are expressed by these three variables ($x$, $\xi$, $t$).
The GPDs of the nucleon,
$H^q$, $E^q$, $\tilde{H}^q$, and $\tilde{E}^q$
are defined by the matrix elements
of nonlocal vector and axial-vector operators as
\vspace{-0.20cm}
\begin{align}
 \int \frac{d y^-}{4\pi}
& e^{i x P^+ y^-}
 \left< p' \left| 
 \bar{q}(-y/2) \gamma^+ q(y/2) 
 \right| p \right>   _{y^+ = \vec y_\perp =0}
\nonumber\\[-0.10cm]
&
=  \frac{1}{2  P^+} \bar{u} (p') 
 \left [ H^q (x,\xi,t) \gamma^+
 +  E^q (x,\xi,t)  \frac{i \sigma^{+ \alpha} \Delta_\alpha}{2 m_N}
 \right ] u (p) ,
\\[-0.10cm]
 \int \frac{d y^-}{4\pi}
& e^{i x P^+ y^-}
 \left< p' \left| 
 \bar{q}(-y/2) \gamma^+ \gamma_5 q(y/2) 
 \right| p \right>  _{y^+ = \vec y_\perp =0}
\nonumber\\[-0.10cm]
&
 =  \frac{1}{2  P^+} \bar{u} (p') 
  \left [ \tilde{H}^q (x,\xi,t) \gamma^+ \gamma_5
     +    \tilde{E}^q (x,\xi,t)  \frac{\gamma_5 \Delta^+}{2 m_N}
 \right ]  u (p) .
\label{eqn:gpd-vector-axial}
\end{align}
\ \vspace{-0.40cm}

\noindent
These GPDs are very useful quantities.
In the forward limit, the GPDs become the unpolarized 
and longitudinally-polarized PDFs.
Their first moments become the corresponding form factors, 
and the second moment is the quark contribution to the nucleon spin:
$J_q  =  \int dx \, x \, [ H^q (x,\xi,t=0) +E^q (x,\xi,t=0) ] /2
      = \Delta q^+ /2 + L_q$.
Here, $L_q$ is a quark orbital-angular-momentum contribution
to the nucleon spin. Because the quark-spin contribution $\Delta q^+$
is known from experimental measurements, it is possible to determine $L_q$
from the GPDs. 
In addition, the GPDs can be used for extracting gravitational 
form factors of hadrons \cite{KSS-grav-2018}, 
so that they are also valuable for finding the origin of hadron masses 
and investigating internal pressure phenomena of hadrons.

\vspace{-0.15cm}
\subsection{GPDs in the ERBL region by $\boldsymbol{2 \to 3}$
process $\boldsymbol{NN \to N \pi B}$}
\vspace{-0.10cm}

\begin{figure}[h!]
\vspace{-0.60cm}
\begin{minipage}{0.55\textwidth}
\vspace{0.60cm}
    \begin{center}
     \includegraphics[width=7.8cm]{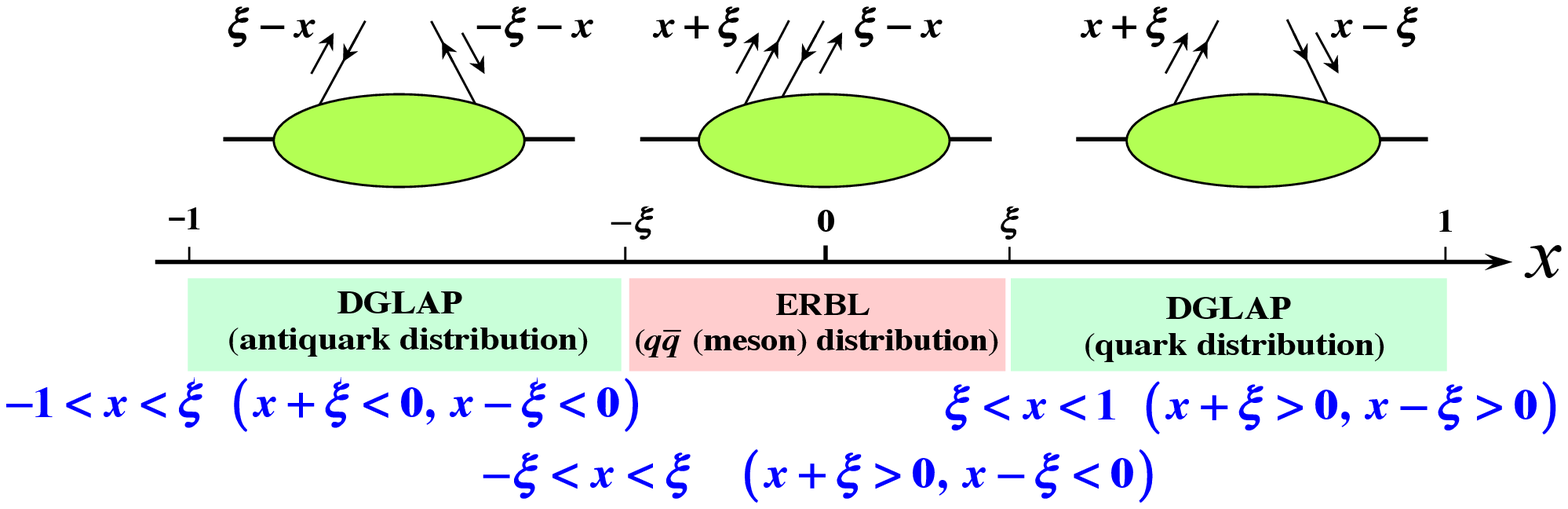}
    \end{center}
\vspace{-0.50cm}
\caption{DGLAP and ERBL regions of GPDs.}
\label{fig:DGLAP-ERBL-GPDs}
\vspace{0.60cm}
\end{minipage}
\hspace{-0.2cm}
\begin{minipage}{0.44\textwidth}
    \begin{center}
     \includegraphics[width=4.5cm]{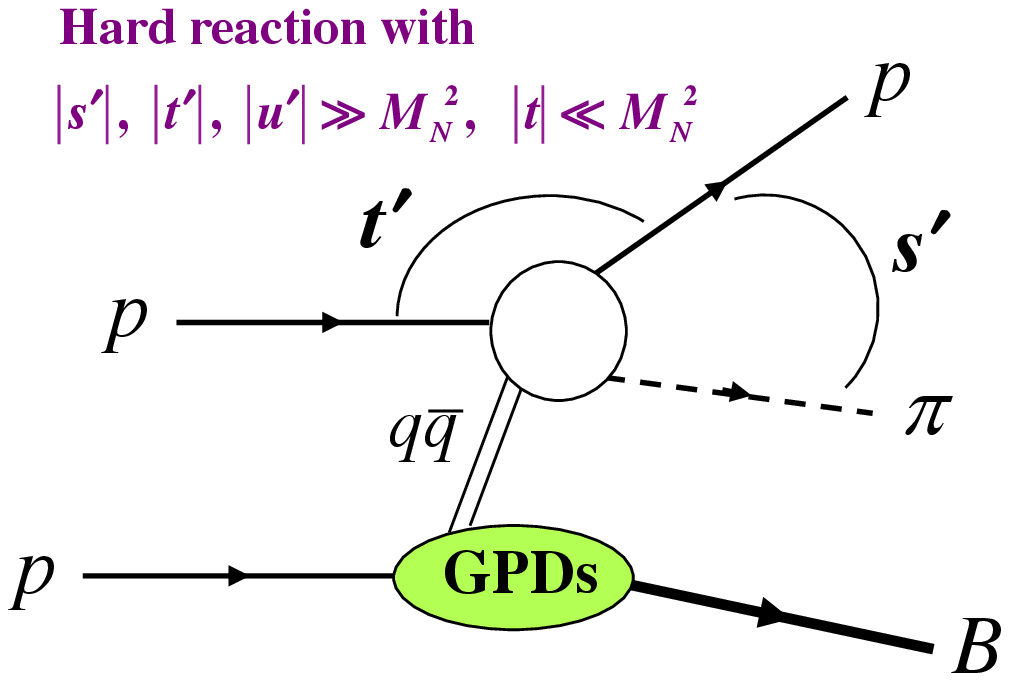}
    \end{center}
\vspace{-0.90cm}
\caption{Typical $2 \to 3$ process $pp \to p \pi B$
and GPD in the ERBL region.}
\label{fig:2-3-GPD}
\end{minipage} 
\vspace{-0.60cm}
\end{figure}

The $2 \to 3$ reaction processes $NN \to N \pi B$ can be used 
for probing the GPDs in a unique kinematical region, which is called
the ERBL (Efremov-Radyushkin-Brodsky-Lepage) region shown in
Fig.\,\ref{fig:DGLAP-ERBL-GPDs}.
The GPDs have three kinematical regions,
(I) $-1<x<-\xi$, (II) $-\xi<x<\xi$, (III) $\xi<x<1$.
The region II is called the ERBL region and the process
corresponds to a quark emission of momentum fraction $x+\xi$
with an antiquark emission of momentum fraction $\xi-x$.
It is a quark-antiquark (meson) distribution amplitude.
The regions I and III are called 
DGLAP (Dokshitzer-Gribov-Lipatov-Altarelli-Parisi) regions,
and the process III (I) corresponds to 
a quark (antiquark) emission 
of momentum fraction $x+\xi$ ($\xi-x$)
with a quark (antiquark) absorption of momentum fraction $x-\xi$ ($-\xi-x$).
It is a quark (antiquark) distribution amplitude.

\begin{wrapfigure}[11]{r}{0.39\textwidth}
   \vspace{-0.20cm}
   \begin{center}
     \includegraphics[width=5.0cm]{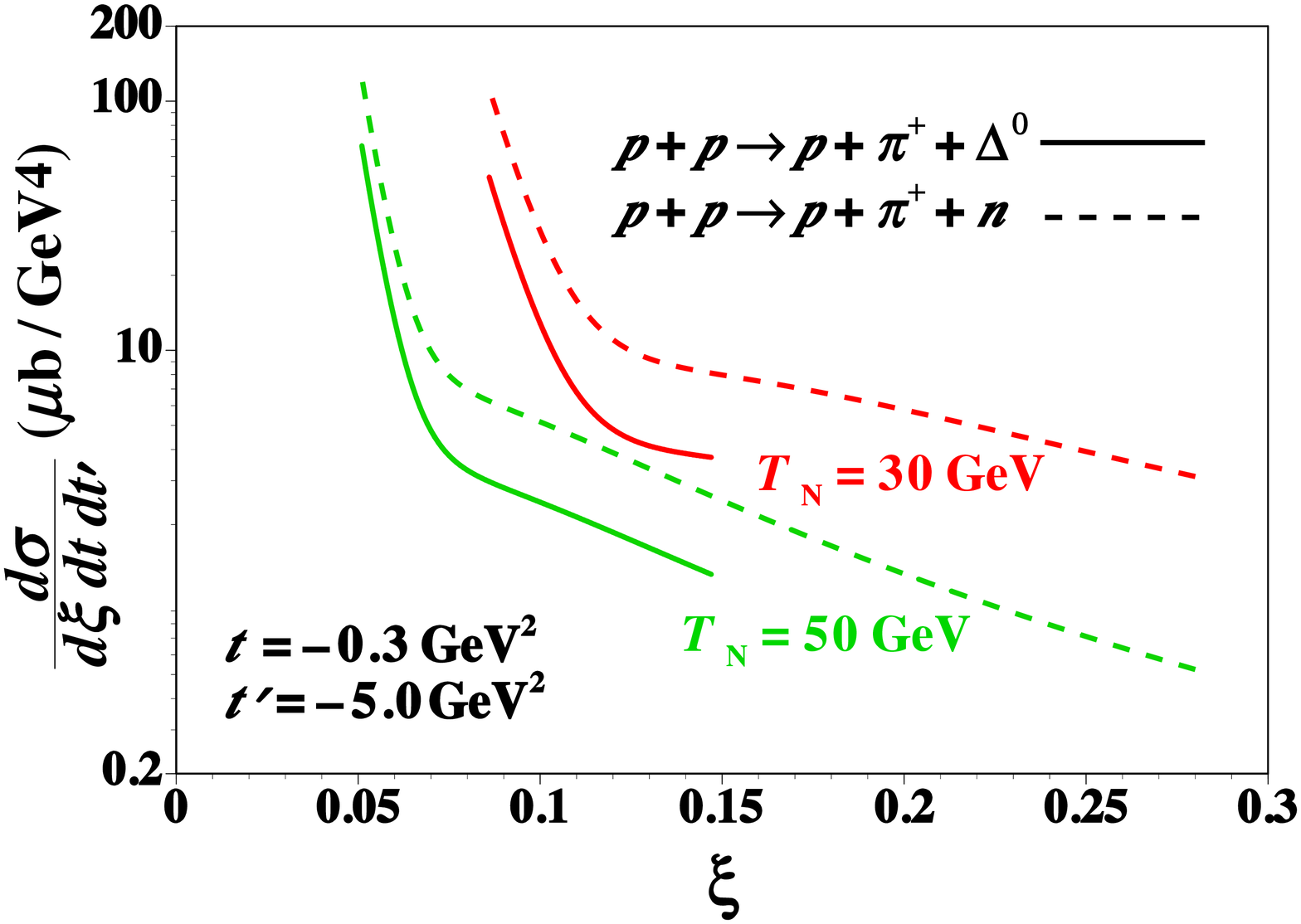}
   \end{center}
\vspace{-0.90cm}
\caption{Cross sections for $pp \to p \pi B$.}
\label{fig:J-PARC-GPD-2-3-cross}
\vspace{-0.5cm}
\end{wrapfigure}

For investigating the ERBL region, we proposed to use the hadronic
$2 \to 3$ process $pp \to p \pi B$ \cite{KSS-GPD-2009}, 
where the baryon $B$ could be a nucleon or $\Delta$, 
in Fig.\,\ref{fig:2-3-GPD}.
If the pion and the final proton 
have large and nearly opposite transverse momenta 
and a large invariant energy, an intermediate exchange 
could be considered as a $q\bar q$ state. This $q\bar q$ 
attached to the nucleon is expressed by the GPDs in the ERBL region.
This experiment is possible at J-PARC, and it is complementary 
to other GPD measurements because of this uniqueness on 
the ERBL kinematics.
The $pp \to p \pi^+ B$ ($B=\Delta^0,\,n$) cross sections are shown
in Fig.\,\ref{fig:J-PARC-GPD-2-3-cross} for the proton beam
energy of 30 and 50 GeV by considering the J-PARC experiment
\cite{KSS-GPD-2009}.
In the hard kinematical region,  
$s', |t'|, |u'|\gg m_N^2$ with $t'/s'={\rm const}$,
the cross section could be factorized into the GPD part
and the hadron ($q\bar q$)-proton scattering part as
$
{\cal M}_{N N \to N\pi B} 
 = {\cal M}_{N \to h B}  \, {\cal M}_{h N\to \pi N}
$,
where $h$ indicates the intermediate $q\bar q$ hadronic state.
Here, the transverse sizes of the projectile and
two outgoing hadrons near the interaction point are
given by $\sim 1/\sqrt{|t'|}$, so that 
the color transparency arguments are essential in proving 
the factorization for an exclusive hadron production
\cite{Collins:1997hv}.
Actually, this factorization was proved for the 
deeply virtual meson-production process in Ref. \cite{Collins:1997hv},
so that we need such a study for the current hadronic reaction.
Depending on the final state $\Delta^0$ or $n$ and the proton-beam energy,
the cross section and the kinematical range vary. 
It is a unique opportunity to investigate the ERBL region of the GPDs.

\subsection{GPDs by exclusive Drell-Yan process $\boldsymbol{\pi^- p \to \mu^+ \mu^- B}$}

\begin{wrapfigure}[11]{r}{0.39\textwidth}
   \vspace{-0.70cm}
   \begin{center}
     \includegraphics[width=5.0cm]{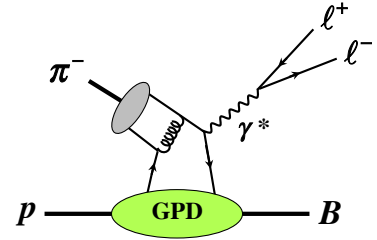}
   \end{center}
\vspace{-0.60cm}
\caption{Exclusive Drell-Yan process.}
\label{fig:J-PARC-GPD-exclusive}
\vspace{-0.5cm}
\end{wrapfigure}

The GPDs can be investigated by the exclusive Drell-Yan process
$\pi^- p \to \mu^+ \mu^- B$ at J-PARC \cite{BDP-2001,J-PARC-GPD-2016}, 
and the experimental proposal is under preparation now
\cite{J-PARC-GPD-LoI}.
This part is discussed by Wen-Chen Chang at this workshop,
so that the detailed should be found in his contribution
\cite{Wen-Chen-Chang-2022}.

The unseparated hadron beam, which is essentially the pion beam,
is available at the high-momentum beamline of J-PARC.
The beam momentum is up to about 20 GeV/$c$.
A typical process is shown in Fig.\,\ref{fig:J-PARC-GPD-exclusive}
with the GPDs.
The nucleonic GPDs are investigated if $B$ is the neutron, 
whereas the transition GPDs are studied for $B \ne N$.
In addition to the GPDs, the pion distribution amplitude
is involved in the reaction; however, 
it could be taken from other theoretical and experimental studies
\cite{pion-distribution}.
Therefore, the GPDs should be extracted from
experimental measurements. For such studies, one should note 
that possibly a large contribution could come from another 
non-factorized process, so called the Feynman mechanism, 
and it needs to be taken into account properly \cite{Tanaka-Feynman-2017}.

\begin{figure}[b!]
\vspace{-1.20cm}
\begin{minipage}{0.55\textwidth}
\vspace{0.60cm}
\begin{center}
     \includegraphics[width=8.5cm]{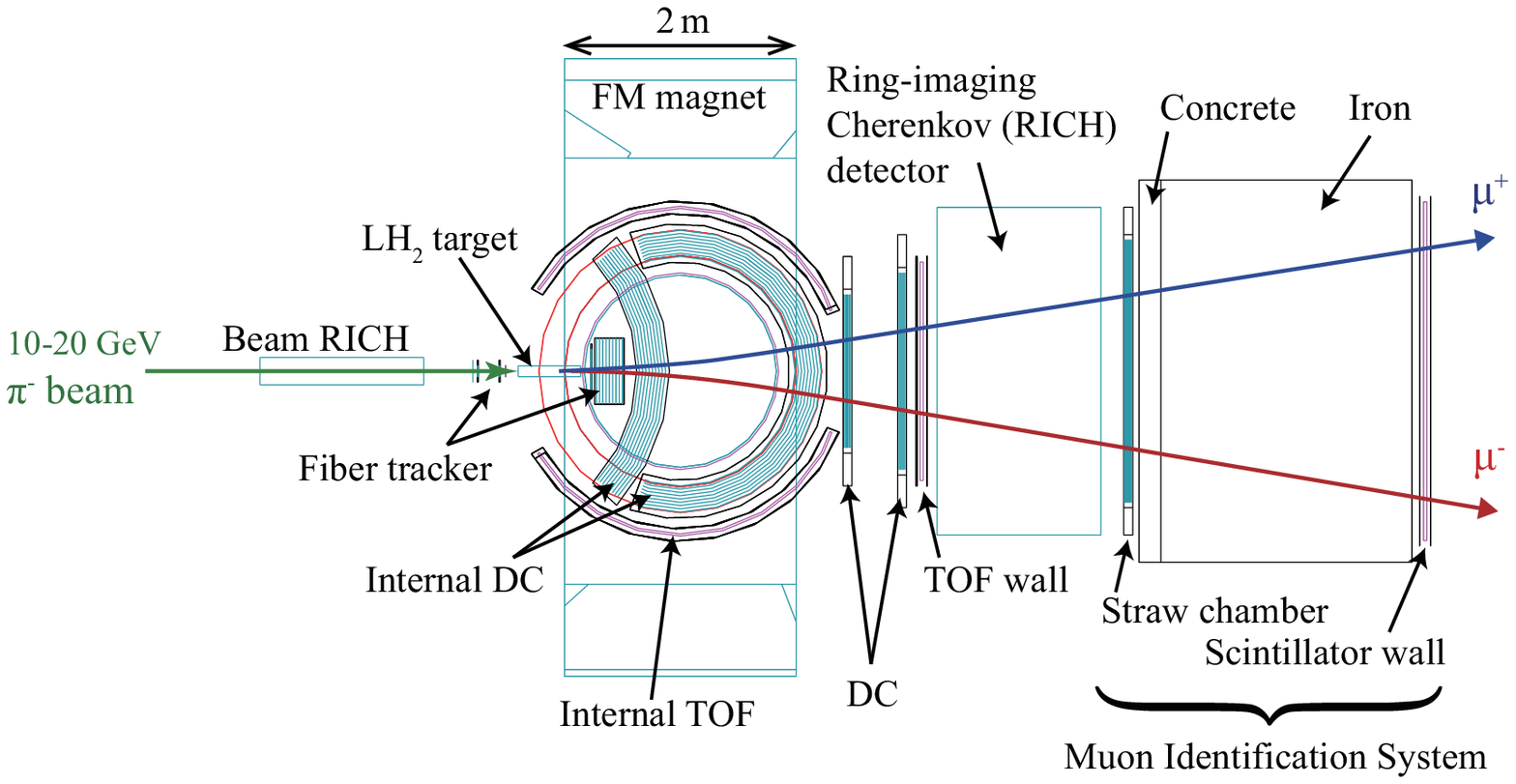}
\end{center}
\vspace{-0.80cm}
\caption{
Detector design for the J-PARC GPD experiment \cite{J-PARC-GPD-2016,APS}.
}
\label{fig:J-PARC-GPD-detector}
\vspace{1.10cm}
\end{minipage}
\hspace{-0.1cm}
\begin{minipage}{0.44\textwidth}
\begin{center}
     \includegraphics[width=4.0cm]{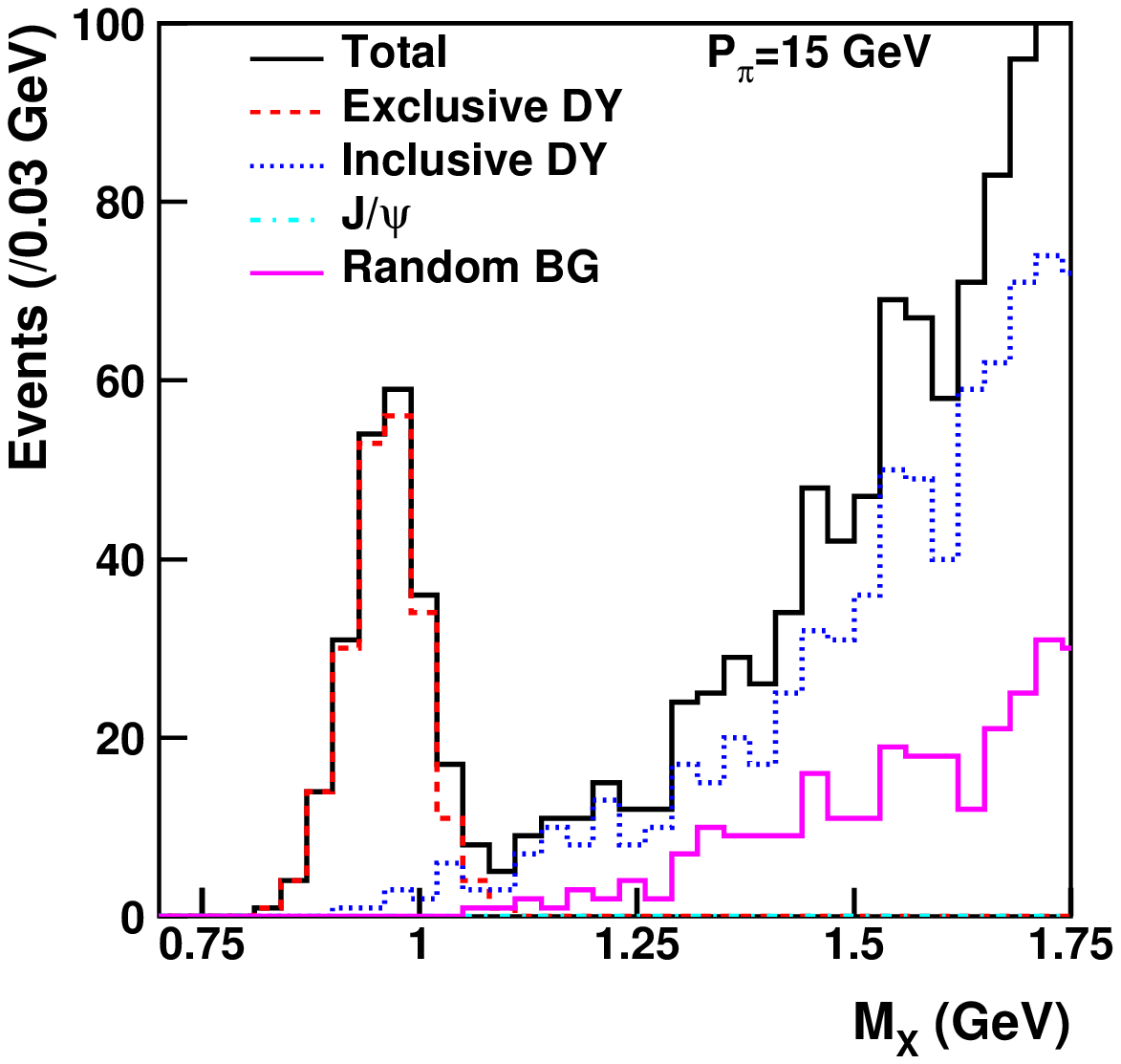}
\end{center}
\vspace{-0.95cm}
\caption{
Monte-Carlo simulation for the missing-mass ($M_X$) spectra
of the $\mu^+ \mu^-$ events at $p_\pi =15$ GeV/$c$
\cite{J-PARC-GPD-2016,APS}.
}
\label{fig:J-PARC-GPD-mx}
\end{minipage} 
\vspace{-0.40cm}
\end{figure}

A new proposal will be submitted soon for this project
\cite{J-PARC-GPD-LoI} in collaboration with the approved E50 experiment
on charmed baryons \cite{J-PARC-E50}. 
In this experiment, the muon identification system is added
to the E50 spectrometer as shown in Fig.\,\ref{fig:J-PARC-GPD-detector}.
The details of the detectors are explained in 
Refs.\,\cite{J-PARC-GPD-2016,J-PARC-GPD-LoI}.
The exclusive Drell-Yan process can be identified by 
the missing-mass technique.
The actual Monte-Carlo simulation is shown in Fig.\,\ref{fig:J-PARC-GPD-mx}
as the function of the missing mass $M_X$ for the pion momentum $p_\pi=15$ GeV/$c$.
It is clear that the exclusive Drell-Yan process is identified
in comparison with other contributions from the inclusive Drell-Yan,
$J/\psi$ production, and random-background processes.
This measurement probes the GPDs in the region $0.1<x<0.3$
by the timelike photon, which is comlementary to the JLab
meson-production experiment with the spacelike photon.
In addition, pion-to-nucleon transition distribution amplitudes
can be investigated by backward charmonium production
in pion-nucleon collisions \cite{TDA-2017}.
Strangeness-hadron production experiments are possible at J-PARC
as mentioned in Ref. \cite{Wen-Chen-Chang-2022},
so that the discussed topics in this paper could be investigated
including strangeness.
Furthermore, we expect that a separated high-momentum kaon beam
will become available in future \cite{j-parc-extension}.
The color-transparency and the $2 \to 3$ GPD 
(Fig.\,\ref{fig:J-PARC-GPD-2-3-cross}) experiments
are still under consideration at J-PARC.
Hopefully, they will be realized in addition to 
the exclusive Drell-Yan experiment.

\section{Summary}

The J-PARC is the most intense proton-accelerator facility
above the multi-GeV region.
By using the high-momentum beams, it is possible to have unique experiments
on the color transparency and the GPDs.
Especially, it should be the key experiment to solve the mysterious 
BNL-AVA result on the color transparency.
In addition, the process $\pi^- A \to \pi^- \pi^+ A^*$ 
($2\to 3$ reaction) provides a unique opportunity to separate 
the transverse size configurations from the space-time evolution, 
in contract to the past $2\to 2$ studies.
Next, the J-PARC GPD experiment by the exclusive Drell-Yan processes
is a complementary project to the JLab one because of 
the smaller-$x$ kinematical region and the timelike photon
instead of the spacelike one.
We also discussed another possible GPD project at J-PARC
by the reaction $pp \to p \pi B$ to investigate
the ERBL region of the GPDs.
All these projects are valuable for the development of
high-energy hadron physics along with other experimental projects
at JLab, Fermilab, KEKB, NICA, FAIR, LHCspin, EIC, EicC, and the others.
From these measurements, we understand basic mechanisms 
of high-energy hadron reactions in nuclear medium.
Furthermore, the GPD studies will lead to clarifications
on the origins of hadron spins, masses, and internal pressures.

\vspace{6pt} 

\section*{Acknowledgments}
S. Kumano was partially supported by 
Japan Society for the Promotion of Science (JSPS) Grants-in-Aid 
for Scientific Research (KAKENHI) Grant Number 19K03830.
The author thanks T. Takahashi for suggestions on the J-PARC experiments.
He also thanks the American Physical Society for 
the copyright permission to use Figures 14 and 15
from Ref.\,\cite{J-PARC-GPD-2016}.

\vspace{-0.10cm}


\begin{thebibliography}{999}
\bibitem{J-PARC-web}
J-PARC web page: 
     https://j-parc.jp/researcher/index-e.html;
Program Advisory Committee for Nuclear and Particle Physics Experiments:
     https://j-parc.jp/researcher/Hadron/en \\ /PAC\_for\_NuclPart\_e.html.
\bibitem{J-PARC-recent-highlights}
Y. Ichikawa {\it et al.}, 
     {\em Prog. Theor. Exp. Phys.} {\bf 2020}, 123D01;
T. Yamaga {\it et al.}, 
     {\em Phys. Rev. C} {\bf 2020}, {\em 102}, 044002;
S. H. Hayakawa {\it et al.}, 
     {\em Phys. Rev. Lett.} {\bf 2021}, {\em 126}, 062501;
Y. Yoshimoto {\it et al.}, 
     {\em Prog. Theor. Exp. Phys.} {\bf 2021}, 073D02.
\bibitem{BNL-EVA-2004}
J. Aclander {\it et al.},
     {\em Phys. Rev. C} {\bf 2004}, {\em 70}, 015208.
\bibitem{jlab-trans-2021}
D. Bhetuwal {\it et al.}, 
     {\em Phys. Rev. Lett.} {\bf 2021}, {\em 126}, 082301.
\bibitem{KSS-GPD-2009}
S. Kumano, M. Strikman, and K. Sudoh,
     {\em Phys. Rev. D} {\bf 2009}, {\em 80}, 074003.
\bibitem{J-PARC-GPD-2016}
T. Sawada, Wen-Chen Chang, S. Kumano, Jen-Chieh Peng, S. Sawada, and K. Tanaka,
    {\em Phys. Rev. D} {\bf 2016}, {\em 93}, 114034.
\bibitem{J-PARC-GPD-LoI}
J. K. Ahn {\it et al.}, 
    {\em Letter of Intent} {\bf 2018}, 
    7th J-PARC PAC meeting, January 16-18, 2019.
\bibitem{Collins:1997hv}
  J. C. Collins, L. Frankfurt, and M. Strikman,
  pp. 296-303, Low x Physics, Proceedings of \\
   the Madrid Workshop, Madrid, Spain, June 18- 21, 1997,
    edited by  F. Barreiro, L Labarga, and \\ J. del Peso, (World Scientific, 1998).
\bibitem{GPD-review}
For introductory review, see 
    M.~Diehl, 
        {\em Phys. Rep.} {\bf 2003}, {\em 388}, 41;  
    X.~Ji,  
        {\em Annu. Rev. Nucl. Part. Sci.} {\bf 2004}, {\em 54}, 413.
For recent works, see
    H. Moutarde, P. Sznajder, and J. Wagner, 
       {\em Eur. Phys. J. C} {\bf 2018} {\em 78}, 890;
    PARTONS project at
       http://partons.cea.fr/partons/doc/html/index.html.
See also Refs. \cite{KSS-GPD-2009,KSS-grav-2018,kk-2014}.
\bibitem{KSS-grav-2018}
S. Kumano, Qin-Tao Song, and O. V. Teryaev,
     {\em Phys. Rev. D} {\bf 2018}, {\em 97}, 014020.
\bibitem{kks-2013}
H. Kawamura, S. Kumano, and T. Sekihara,
     {\em Phys. Rev. D} {\bf 2013}, {\em 88}, 034010.
\bibitem{kk-2014}
H. Kawamura and S. Kumano, 
     {\em Phys. Rev. D} {\bf 2014}, {\em 89}, 054007.
\bibitem{cks-2016}
Wen-Chen Chang, S. Kumano, and T. Sekihara, 
     {\em Phys. Rev. D} {\bf 2016}, {\em 93}, 034006.
\bibitem{ss-2011}
R. A. Schumacher and M. M. Sargsian,
{\em Phys. Rev. C} {\bf 2011}, {\em 83}, 025207.
\bibitem{abrs-2021}
M. J. Amaryan, W. J. Briscoe, M. G. Ryskin, and I. I. Strakovsky,
{\em Phys. Rev. C} {\bf 2021}, {\em 103}, 055203.
\bibitem{JLab-counting-2003}
L. Y. Zhu {\it et al.},  
     {\em Phys. Rev. Lett.} {\bf 2003}, {\em 91}, 022003.
\bibitem{2010-ks}
S. Kumano and M. Strikman,
     {\em Phys. Lett. B} {\bf 2010}, {\em 683}, 259.
\bibitem{GPD-neutrino}
S. Kumano and R. Petti,
     {\em PoS (Proceedings of Science)}, {\bf 2022}, {\em NuFact2021}, to be published.
\bibitem{BDP-2001}
  E. R. Berger, M. Diehl, and B. Pire, 
  {\em Phys. Lett. B} {\bf 2001}, {\em 523}, 265.
\bibitem{Wen-Chen-Chang-2022}
  Wen-Chen Chang, contribution to this workshop 
  at https://indico.jlab.org/event/437/.
\bibitem{pion-distribution}
For the recent situation of the pion distribution amplitude, see
    C. D. Roberts,  D. G. Richards, T. Horn, and L. Chang,
{\em Prog. Part. Nucl. Phys.} {\bf 2021}, {\em 120}, 103883. 
\bibitem{Tanaka-Feynman-2017}  
  K. Tanaka, 
  arXiv:1703.02190, contribution to the Spin2016 symposium;
  K. Tanaka, \\
  {\em PoS (Proceedings of Science)} {\bf 2017} {\em DIS2017}, 249.      
\bibitem{J-PARC-E50}
KEK/J-PARC-PAC 2012-19, 
P50 proposal, Charmed Baryon Spectroscopy via the ($\pi$,$D^{*-}$) reaction,
http://www.j-parc.jp/researcher/Hadron/en/Proposal\_e.html,
see also https: \\ //www.rcnp.osaka-u.ac.jp/$\sim$noumi/puki/E50/.
\bibitem{APS}  
Figures 14 and 15 are used with the copyright permission of the American
Physical Society and authors from Ref.\,\cite{J-PARC-GPD-2016}.
\bibitem{TDA-2017}
 B. Pire, K. Semenov-Tian-Shansky, and L. Szymanowski,
     {\em Phys. Rev. D} {\bf 2017}, {\em 95}, 034021.
\bibitem{j-parc-extension}
For the future hadron-hall extension at J-PARC, see
the J-PARC workshop on Review of Hadron Experimental-Facility Extension at
https://kds.kek.jp/event/38930/ .
\end{thebibliography}
\end{document}